\documentstyle[12pt,doublespace]{article}
\def\phys{Phys. Rev. B}
\def\lett{Phys. Rev. Lett.}
\def\riera{Riera$^1$ and Elbio Dagotto$^2$}
\def\cente{$^1$Center for Computationally Intensive Physics}
\def\depte{$^2$Department of Physics}

\begin{document}
\title{\protect\Large \bf Superconductivity in the CuO Hubbard model
with long-range Coulomb repulsion}

\author{Jos\'{e} \riera   \\
\vspace{-3 mm}
\protect\small \em \cente ,
\protect\small \em Physics Division,             \\
\vspace{-3 mm}
\protect\small \em Oak Ridge National Laboratory, Oak Ridge, Tennessee 37831,
\\
\vspace{-3 mm}
\protect\small \em and Department of Physics $\&$ Astronomy, Vanderbilt
Universi
ty,           \\
\vspace{-3 mm}
\protect\small \em Nashville, Tennessee 37235. \\
\vspace{-3 mm}
\protect\small \em \depte ,
\protect\small \em National High Magnetic Field Laboratory,     \\
\vspace{-3 mm}
\protect\small \em and MARTECH, Florida State University,
Tallahassee, Florida 32306.}
\date{October 1993}
\maketitle

\vspace{-5 mm}
\begin{abstract}
A multiband CuO Hubbard model is studied which incorporates long-range (LR)
repulsive Coulomb interactions. In the atomic limit, it is shown that
a charge-transfer from copper to oxygen ions occurs as the
strength of the LR interaction is increased. The regime of phase
separation becomes unstable, and is replaced by a uniform state with doubly
occupied oxygens. As the holes become mobile a superfluid condensate
is formed, as suggested by a
numerical analysis
of pairing correlation functions and flux quantization. Although most
of the calculations are carried out on one dimensional chains, it is
argued that the results are also applicable to two dimensions.
\end{abstract}

PACS Indices: 75.10.Jm, 75.40.Mg, 74.20.-z\hfill

\vspace{0.4 cm}

\newpage

The study of high-T$_c$ superconductors$^1$ continues
attracting much attention.
Recent
progress has been made in the theoretical search for ground state
superconductivity in one band electronic models. The two dimensional
${\rm t-J}$ model has shown indications of d-wave superconductivity$^2$
when analyzed near the phase separation regime$^3$ at intermediate
densities.
It would be desirable to extend these observations about
superconductivity to more
realistic three band versions of the Hubbard model, such as those
introduced by Emery$^4$ and Varma and
collaborators.$^5$
Not much is known for the superconducting properties of these models.
Weak coupling and large N mean-field
calculations applied to this
extended model have indeed shown that a region of superconductivity
exists near phase separation as in the one band
case.$^6$
However, these techniques are only approximate and should be
supplemented by studies of the multiband Hubbard
model with unbiased computational methods, like exact diagonalization
and quantum Monte Carlo$^{7,8}$.

Varma and collaborators$^5$ have suggested that the inclusion of
a short-range Coulomb repulsion induces a charge transfer process,
leading to
the formation of tightly bound hole pairs on the oxygen ions. If these pairs
are mobile they naturally lead to a superfluid condensate.
Unfortunately, the charge transfer mechanism seems systematically
correlated with a phase separation process, as was recently discussed in
the atomic limit.$^9$ Phase separation occurs between a phase
with a density of one hole per unit cell on the copper ions and a
second
phase which has a density of two holes per unit cell, with the charge
located on double occupied oxygens. These pairs are not mobile, and thus
the system cannot become superfluid. In this sense, phase separation is
an unwelcomed effect in this model.
In spite of this problem, Sudb\o$\:$
et al.$^{10}$ recently
found indications that the one dimensional
model exhibits superconducting correlations immediately before
phase separation. This conclusion is based on a study
of the conformal field theory parameter $K_{\rho}$; $K_{\rho} > 1$
indicates that superconducting power-law correlations are dominant in the
ground state. (Unfortunately the actual pairing correlations in the ground
state decay rapidly with distance$^{11}$.)
While this result is very encouraging, it is difficult to predict in
advance whether a region of superconductivity would exist near phase
separation in the more realistic two dimensional
problem.
Thus, it would be desirable to have a model wich exhibites
robust superconducting correlations in the ground state,
and whose main features can be understood intuitively, allowing its
extension to two dimensions where explicit numerical simulations are
difficult.

In this paper, a simple modification of the standard three band
model is presented that addresses these issues. Even
in the atomic limit, a clear separation between the parameter values
associated with charge transfer
and phase separation is evident. Including a hopping term,
robust numerical evidence of superconducting correlations is observed.
This three band Hubbard model for the Cu-O cuprates is defined by,$^{4,5}$
\begin{eqnarray}
H = - t_{pd} \sum_{\scriptstyle <i j> \sigma}
(c^{\dagger}_{i \sigma}c_{j \sigma} +
c^{\dagger}_{j \sigma}c_{i \sigma})
+ U_d \sum_{i}  n_{i \uparrow}n_{i \downarrow}
+ U_p \sum_{j}  n_{j \uparrow}n_{j \downarrow}
+ \epsilon_d \sum_{i}  n_i
+ \epsilon_p \sum_{j}  n_j,
\end{eqnarray}
\noindent
where the operators are hole number and hole creation and annihilation
operators,
$U_d$ and $U_p$ are the Coulomb on-site repulsion energies
for the copper and
oxygen sites respectively, and $\epsilon_d$ and $\epsilon_p$ are
the ion energies.
The label $i$ corresponds to Cu sites, and $j$ denotes O sites
and they
are connected by a hopping amplitude $t_{pd}$.
The charge transfer energy is defined by  $\Delta = \epsilon_p -
\epsilon_d$.
It is generally accepted that $U_d > \Delta, U_p$ in the
cuprates, so we will work in this regime.
The doping fraction is ${\rm x = n_h/N}$,
where ${\rm n_h = (N_h-N)}$ is the number of doped holes,
${\rm N_h}$ is the total number of holes and
${\rm N}$ is the number of
Cu-O cells.
At half-filling, ${\rm N_h = N}$, as in the insulating Cu-O
compounds. Most of the results presented here correspond to
low and intermediate doping ($x \leq 0.66$).

As explained before,
Varma et al.$^5$ argued that the Hamiltonian Eq.(1) needs to
be supplemented by
a nearest-neighbor Coulomb repulsion,
\begin{eqnarray}
H_{nn} =  V \sum_{\scriptstyle <i j>}  n_{i}n_{j} ,
\end{eqnarray}
\noindent
(where V is a positive number, and $n_{i,j}$ are hole number operators)
to induce the charge transfer
mechanism, but this term also induces
phase separation, at least in the atomic limit.
To avoid this problem we have extended the range of
the density-density correlations of Eq.(2).
The relevance  of the long range part of the Coulomb
interaction between the holes, which prohibites phase separation
was emphasized in a recent work by Emery and Kivelson.$^{12}$
In this work, experimental evidence was provided in support of
the ``frustrated phase separation" scenario which results in
the presence of long range Coulomb interaction.$^{13}$
In particular, if the interactions are
of infinite range, then phase separation is certainly eliminated.$^{14}$
This leads us to replace Eq.(2) by a more general interaction
\begin{eqnarray}
H_C = {{V}\over{2}} \sum_{\scriptstyle l \neq m}
\frac{n_l n_m}{d_{lm}} e^{ - \frac{d_{lm}}{\lambda}},
\end{eqnarray}
\noindent
where $d_{lm} = |{\bf r}_l - {\bf r}_m |$, and $\lambda$ is the range of the
screened Coulomb repulsion. If
$\lambda = \infty$, the interaction decays
as the inverse of the distance, and it is of infinite range.
Most of the results presented are for this limit.
We also consider finite values of $\lambda$ to show that our
conclusions are qualitatively valid even for
finite range forces, provided that they operate over distances larger
than a few lattice spacings.

To motivate the introduction of a long-range term in
the Hamiltonian it is convenient to first consider the
atomic limit $t_{pd} = 0$.
Let us consider the one dimensional case first. In this case, it
is not difficult to guess which are the possible states with lowest energy
for different values of V. We have then checked that some of these
states are effectively the ground state in a certain region of V
using a simulated annealing algorithm described below.
In Fig.1a we show the ground state in the regime of small V.
This state (denoted by I),
has a hole at every copper ion, and the remaining of
the holes are on singly occupied oxygens, distributed in a regular pattern to
minimize the newly introduced long-range potential energy.
In this region, for the two dimensional (2D) model, and for not too
high doping, some antiferromagnetic (AF) correlations are expected
to develope.
On increasing V a phase separated state (denoted by II and
shown in Fig.1b) becomes energetically competitive with state I.$^{9}$
It contains a large region of doubly occupied oxygens
and another separated region with
one hole per copper ion. These two densities coexist and changing
the number of holes just increases or reduces their relative size.
However, there is a third possibility which becomes energetically
favorable due to the long-range Coulomb interaction. In Fig.1c we show
a state (III) where all the holes are on doubly occupied
oxygens.
The reason for considering this state is that
charge tends to spread uniformly over the lattice
in the presence of long-range forces, rather than
separating into different densities, as in the phase separated
state. State III is a configuration with ``preformed''
pairs on the oxygens, and if this state becomes stable in the
atomic limit, a finite hopping amplitude may render it superfluid.
In other words, once the pairs are preformed at zero hopping, it is
reasonable to expect that they will behave as hard-core bosons when
they are allowed to move, at least for a small hopping amplitude.
Similar behavior is observed in the
attractive Hubbard model at large ${\rm |U|/t }$. Note that state III resembles
a Wigner crystal of charge 2e pairs at the oxygens which is precisely
expected to be stable due to long-range interactions.
We have compared the energies of states I, II and III
numerically in the atomic limit for 12, 24 and 48 CuO cells.
We have considered $U_d = 7$, $U_p = 1$, $\Delta = 1.5$, in
arbitrary units, and $x = 0.33$. For the 12 cell lattice there
is a small interval of V, [3.1, 3.7], where state II has the
lowest energy. For the 24 cell lattice this interval is smaller
and finally, for the 48 cell lattice there is a direct crossing from
state I to state III at $V_c \approx 3.38$.
Similar arguments are  obviously valid in the two
dimensional case, so our results are not restricted to chains
only.
For illustration we have carried out
an explicit calculation for a $4 \times 4$ CuO$_2$ periodic cluster
with four doped holes ($ x = 0.25$).
In this case the ground state is not obvious, and it is difficult
to determine it by calculating the energy of each possible configurations
because their number grow exponentially with the number of atoms.
For this reason we developed a simulated annealing$^{15}$ program to find
the ground state. This algorithm can be described as a sequence of
Markov chains, each one generated at a fixed value of the temperature.
The temperature is decreased between subsequent Markov chains.
At each step of the Markov chain, the algorithm attempts to transform
the current state
into an state obtained by moving one hole to a nearest-neighbor site.
The variation of the energy is computed,
and the local change is accepted according to the Metropolis criterion.
Using this technique we have determined the ground state of this cluster
for the same set of parameters U$_d$, U$_p$ and $\Delta$ as before.
For small V, the ground state has one hole per Cu atom and the doped
holes are in O sites locates as far as possible from each other.
As $V$ is increased, a charge transfer from Cu to O sites starts to
develope.
For $V \approx 3$, a state with doubly occupied O sites
separated from a region of single-occupied Cu sites becomes
stable. For larger values of $V$, the ground state has the form
illustrated in Fig.1d.
These results in two dimensions give further support to
our intuitive conclusion that in the presence of strong
long-range interactions, there is a phase with preformed pairs
without phase separation.

In order to find the $\lambda$ dependence of our results,
numerical calculations where carried out using the simulated annealing
technique on chains of 12 and 24 CuO cells ($x = 0.33$).
We have considered the same parameters as above.
As an order parameter we used the number of doubly occupied
oxygen sites.
The results are shown in Fig.2a. We observed that
our intuitive ideas about states I, II, and III are qualitatively correct,
and they are dominant in a large region of parameter space.
Only in narrow regions near the phase boundaries, especially for large
values of $\lambda$, did we observe
other types of states become
competitive. However, these states are simple variations
of states I, II, and III (for example, including small fluctuations),
so the essential features shown in Fig.2a are correct.
Note that, for the 12 cell chain, there is a particular value
of $\lambda \sim 8$ where the
phase separated regime becomes very narrow, and it may
dissappear when the hopping parameter is switched on. (State III
will reduce its energy by providing mobility to the pairs, but state
II is too rigid to take advantage of this possibility.)
This value of $lambda$ reduces to $\sim 4$ for the 24 cell chain.
We conjecture
therefore that the phase separated regime does not exist for
$\lambda > 4$ or perhaps for even smaller values of $\lambda$, so
it is not necessary to have a fully unscreened 1/r interaction to obtain the
effects described here. This
value of $\lambda$ corresponds to two Cu-Cu lattice spacings (about $7.6
\AA$ in the cuprates),
and thus it is not
physically unrealistic.
The main features in Fig.2a are also present for the case
of 8 doped holes ($x = 0.66$). Similar calculations for the
$4 \times 4$ and $6 \times 6$ CuO$_2$ clusters, for $x \approx 0.25$,
give essentially
the same phase diagram, as it can be seen in Fig.2b.

The results summarized in Figs.1 and 2a,b were obtained in the atomic
limit. Although they are very suggestive, the ultimate test of these ideas
requires a finite kinetic energy term to give mobility to the carriers.
The actual boundaries in parameter space between different ground
states will be determined by the competition between the Coulomb
potential energy and the kinetic energy. To study this interplay, and
to establish the assumed existence of a superconducting regime, Lanczos
diagonalization techniques were used to obtain the ground state of
the Hamiltonian Eq.(1,3) explicitly on a finite chain. Due to the
large number of states per unit cell the lattices that can be studied
numerically are limited to a small number of
cells. In this paper we
studied 6 cells (12 atoms) with various hole numbers. For the
results shown below we chose $U_d = 7$, $U_p = 1$ and $\Delta =
1.5$ in units of $t_{pd}$.
These are reasonably realistic values
for the cuprates.$^{16}$ We have also observed that our results are
qualitatively similar over a broad range of parameters.

First, let us numerically consider the issue of charge-transfer versus phase
separation numerically.
A quantitative measure of charge transfer from Cu to O sites
is given by the expected occupation of oxygen sites. However,
a sharper indication of charge transfer
is given by the susceptibility
associated with that quantity, $\chi_{CT} = < n_O^2 > - < n_O >^2$,
($n_O$ is the number of holes on the oxygens).$^8$
The results for this susceptibility are shown in Fig.2c for
$n_h = 2$ and $n_h = 4$ on the 12 atom chain and a long range
interaction.
In both cases a peak is observed at a particular value of
$V \approx 4$. This peak presumably signals the onset of a
charge transfer process in this model. (Note also that $\chi_{CT}$ in the
region ${\rm V > U_d}$ shows additional structure, although this is an
unphysical region.)
$\chi_{CT}$ however does not
distinguish between the states II and III in Fig.1.
To confirm that phase separation does not occur in the presence of
long-range interactions, we have measured the short wavelength
component of the susceptibility associated with the correlations
of pairs of holes on O sites.$^{17}$ This quantity (denoted by X)
is normalized to 1 for a fully phase
separated state. In Fig.2d, X is shown for the case
$n_h = 2$ on the 12 atom chain. With a short range interaction,
${\rm X}
\rightarrow 1$ as V is increased. On the other hand, in the presence
of long-range
interactions, X goes through a maximum, and then converges to zero
at large V, showing the absence of phase separation in this limit.
The peak at intermediate V is due to the proximity of
the phase separated state in the spectrum (as discussed previously in the
atomic limit). At this point it is important to remark that the
critical value of V (${\rm V_c}$) at which the charge transfer process occurs
(namely
${\rm \sim 3 t_{pd}}$) is physically acceptable.
Some calculations suggest that
${\rm t_{pd} \sim 1.3 eV}$ in the cuprates,$^{16}$ and thus ${\rm V_c}$
obtained here is approximately 4eV. This is not far from
estimations$^{18}$ of V based on measurements of the dielectric constant that
suggest ${\rm V \sim 2 - 3 eV}$.

Now let us investigate whether state III of Fig.1c
becomes superfluid when the holes
acquire mobility in the presence of LR interactions.
For this purpose, we consider the pairing correlation,
\begin{eqnarray}
C(m) = \frac{1}{N} \sum_{\scriptstyle j} < \Delta_{j+m}^{\dagger}
\Delta_{j} >,
\end{eqnarray}
\noindent
where the pairing operator is defined as
$\Delta_j = c_{j \uparrow} c_{j \downarrow}$, and
$j, j+m$ indicates O sites.
In Fig.3a-b, the pairing correlations are plotted as a function of
distance for two different densities, for $t_{pd} = 1$, $U_d =
7$, $U_p = 1$, and $\Delta = 1.5$, and several values of $V$. For ${\rm
n_h = 2}$
the correlations are robust at distance one which is the distance
between two oxygen sites ($ \sim 3.8 \AA$ in the cuprates). At larger
distances
the correlations decay rather rapidly, so our hypothesis is not
clearly supported by these results (note however that these correlations
are already
stronger than those found in one band models near half-filling).
On the other hand, with 16 holes on the chain ($ n_h = 4$)
the pairing correlation is very
strong even at the largest distances available on our finite clusters.
The signal monotonically increases with V, for $V < U_d$.$^{19}$

The results presented thus far support a scenario in which
s-wave (local) superconductivity
appears with increasing V if LR interactions are included.
To give further support to these claims,
we have studied the dependence of the ground state energy on
an external magnetic flux $\Phi$. To analyze this
response, a phase factor $e^{i \Phi/N}$ is introduced in the
hopping term in Eq.(1).$^{2,10}$  This is equivalent
to allowing a magnetic flux through the Cu-O ring (chain with periodic
boundary conditions).
In Fig.4,
the ground state energy measured with
respect to the energy at zero flux, $\Delta E = E(\Phi) - E(\Phi = 0)$,
is shown as a function of $\Phi$ for  $n_h = 4$ and a pure
LR interaction ($\lambda = \infty$).
For small $V$ the energy has a single minimum
at $\Phi = 0$ $(mod 2\pi)$.
For $V$ larger than a critical value
the energy develops a second
minimum at $\Phi = \pi$,
indicating the presence of mobile carriers with charge $2e$ in the
ground state.$^{20}$
This anomalous flux quantization is in agreement
with results obtained from the pair correlations.

In summary, we have considered an extension of the standard Cu-O electronic
Hubbard-like model for the
cuprates which incorporates long-range interactions. We
found that in the atomic limit a charge-transfer process exists without
phase separation. The new dominant state in this regime
has preformed pairs on the oxygen sites. With a hopping term that provides
mobility to the
holes, this state appears to become superfluid. The effects observed
here are intuitively evident, and are valid both in one and
two dimensions. These effects occur not only with a pure 1/r
interaction but also for a finite-range Coulomb repulsion
($\lambda \sim 4$) which might correspond to the
real cuprate compounds.

The authors thank V. Emery, A. Moreo, L. Testardi,
and A. Aligia for useful conversations.
J. R. was supported in part by the U. S. Department of Energy (DOE)
Office of Scientific Computing under the High Performance Computing and
Communications Program (HPCC), and by DOE under contract No.
DE-AC05-84OR21400 managed by Martin Marietta Energy Systems, Inc., and
contract No. DE-FG05-87ER40376 with Vanderbilt University.
E. D. was supported by the Office of Naval Research under
grant
ONR N00014-93-1-0495.
We thank the
Supercomputer Computations Research Institute (SCRI) for its support.
Some of the numerical work was done on the CRAY-YMP at the
National Center for Supercomputing Applications, Urbana, Illinois.

\newpage

\vspace{2cm}
\centerline{\large \bf References}

\begin{enumerate}

\singlespace

\item J. Bednorz and K. M\"uller, Z. Phys. {\bf B 64}, 188
(1986); Rev. Mod. Phys. {\bf 60}, 585 (1988).

\item E. Dagotto and J. Riera, \lett {\bf 60}, 682, (1993).

\item
V. J. Emery, S. A. Kivelson and H. Q. Lin, \lett {\bf 64},
475 (1990).

\item V. J. Emery, \lett {\bf 58}, 2794 (1987).

\item
C. M. Varma,
S. Schmitt-Rink, and E. Abrahams, Solid State Comm. {\bf
62}, 681 (1987).

\item P. B. Littlewood, C. M. Varma, and E. Abrahams,
\lett {\bf 63}, 2602 (1989);
P. B. Littlewood et al., \phys {\bf 39}, 12371 (1989);
M. Grilli, R. Raimondi, C. Castellani, C. Di Castro, and G. Kotliar,
\lett {\bf 67}, 259 (1991); C. Di Castro and M. Grilli, Physica Scripta
Vol. T45, 81 (1992);
R. Raimondi, et al., \phys {\bf 47}, 3331 (1993).

\item For a recent review see E. Dagotto, ``Correlated Electrons in High
Temperature Superconductors'', NHMFL preprint (1993), to appear in Rev.
Mod. Physics.

\item
R. T. Scalettar, D. J. Scalapino, R. L. Sugar, and S. R.
White, \phys {\bf 44}, 770 (1991).
See also M. Frick, P. Pattnaik, I. Morgenstern, D. Newns, and W. von
der Linden, \phys {\bf 42}, 2665 (1990); G. Dopf, A.
Muramatsu, and W. Hanke, \phys {\bf 41}, 9264 (1990).

\item A. Sudb\o, S. Schmitt-Rink, and C. M. Varma, \phys
{\bf 46}, 5548 (1992).

\item
A. Sudb\o, C. M. Varma, T. Giamarchi, E. B. Stechel, and R. T.
Scalettar, \lett {\bf 70}, 978 (1993).
See also K. Sano and Y. Ono, Physica {\bf C 205}, 170 (1993).

\item E. Dagotto, J. Riera, Y. C. Chen, A. Moreo, A. Nazarenko,
F. Alcaraz, and F. Ortolani, FSU preprint, Sept. 1993.

\item
V. J. Emery, and S. A. Kivelson, Physica {\bf C 209}, 597 (1993).

\item
Experimental results suggesting the presence of phase separation
in $La_2 Cu O_{4 + \delta}$ have been presented in
P. C. Hammel, et al., \phys {\bf 42}, 6781 (1990);
Physica {\bf C 185}, 1095 (1991).

\item Previous work in the three band Hubbard model and
t-J models have also found that phase separation is destroyed by
a long-range interaction. For example, see
G. A. Medina and M. D. N\'unez Regueiro, \phys {\bf B42},
8073 (1990); A. Aligia, H. Bonadeo, and J. Garces, \phys {\bf B 43}, 542
(1991); and
M. Troyer, H. Tsunetsugu, T. M. Rice, J. Riera, and E. Dagotto,
\phys {\bf 48}, 4002 (1993).

\item S. Kirkpatrick, J. of Stat. Phys. {\bf 34}, 975 (1984).

\item
See for example, M. S. Hybertsen, E. B. Stechel, M. Schluter and
D. R. Jennnison, \phys {\bf B 41}, 11068 (1990); E. B. Stechel
and D. R. Jennison, \phys {\bf B 38}, 4632 (1988).

\item C. M. Varma, private communication.

\item F. F. Assaad and D. Wurtz, \phys {\bf 44}, 2681 (1991).

\item In the atomic limit, for ${\rm V > U_d}$, another state in
which there are
both Cu and O doubly occupied sites becomes energetically
favourable. This state is more rigid than the state III, so
the signal slowly decreases with increasing V.

\item Note that in 1D there is no Meissner effect, and thus
the superfluid density cannot be obtained from the curvature of the
energy versus flux, at zero flux. This quantity only measures the
Drude weight contribution to the optical conductivity.

\end{enumerate}

\newpage

\singlespace

\vspace{2cm}
\centerline{\large \bf Figure Captions}
\vspace{1cm}

\begin{description}

\item [FIG. 1] a) Ground state of the Hamiltonian Eq.(1) with long-range
interactions for small V, in the atomic limit.
The copper ions are represented
by large circles and the oxygens by small circles. In this and the
following figures in the atomic limit the alignment of spins is
arbitrary. This state is ground state in the large $U_d$ regime if
$U_p$ is nonzero, otherwise it becomes energetically favorable to move
the holes of singly occupied oxygens into
doubly occupied oxygens.
b) Ground state of the model in the region of small $\lambda$
(short-range interactions) and large V. This state has phase separation
and corresponds to the state previously discussed in Ref.[9];
c) New state discussed in this paper. It becomes the ground state of
the problem in the case of large $\lambda$ (i.e. with long-range
interactions) and sufficiently large V. This state has preformed pairs,
and is not phase separated.
d) Generalization of the state shown in Fig.1c to a two dimensional
lattice.

\item [FIG. 2] a) Phase diagram of the Hamiltonian Eqs.(1,3) in the atomic
limit, obtained from a chain with 12 CuO cells (solid lines) and from
a chain with 24 CuO cells (dotted lines), $U_d = 7$, $U_p = 1$,
and $\Delta = 1.5$ (in arbitrary units), as a function of $\lambda$
and $V$. The doping fraction is $x = 0.33$. The boundaries
were obtained numerically using a simulated annealing algorithm. ``PM''
denotes a region dominated by the states like the one shown in
Fig.1a; ``Ph.Sep.'' denotes the region where
state Fig.1b dominates; and ``Pairs'' is the region with
preformed
pairs in the ground state, as exemplified by Fig.1c and 1d.
b) Same as a) for the $4 \times 4$ (solid lines) and $6 \times 6$
(dotted lines) and $x = 0.25$.
c) Susceptibility $\chi_{CT}$ (see text) as a function of
V for a 6 cell chain (12 atoms),
$t_{pd} = 1$ and other parameters as in Fig.2a.
The full squares denote $n_h = 2$ and the circles $n_h = 4$.
d) Order parameter X (see text) used to study phase separation
versus V.
The full squares
denote $\lambda = \infty$, and the open circles are for the
nearest-neighbor interaction. The number of holes is 8 ($n_h = 2$).

\item [FIG. 3] a) Superconducting correlations ${\rm C(m)}$
(as defined in the text) for the long-range model at several
values of V, on a chain with 12 atoms (6 cells) and
8 holes (${\rm n_h = 2}$). Note that $m$ measures O-O distances.
The full triangles
denote results for ${\rm V = 2 }$, the open squares ${\rm V = 4}$,
and the full squares ${\rm V = 6}$;
b) As a) but with 10 holes (${\rm n_h = 4}$).

\item [FIG. 4] The ground state energy $\Delta E(\Phi)$
(measured with respect to the energy at zero flux)
versus flux $\Phi$ for a 6 cell-ring
for various V and a long-range interaction ($\lambda = \infty$).

\end{description}

\end{document}